\documentclass[article]{aa}
\usepackage{graphicx}
\usepackage{txfonts}

\usepackage{natbib}

\usepackage[english]{babel}

\begin{document}
\title{Signature of [SiPAH]$^+$ $\pi$-Complexes in the Interstellar Medium
\thanks{This work is based on observations made with ISO, an ESA project with instruments funded
by ESA Member States (especially the PI countries: France, 
Germany, the Netherlands and the United Kingdom) and with the
participation of ISAS and NASA.}}

\author{B. Joalland \inst{1,2}
\and 
A. Simon \inst{1}
\and
C. J. Marsden \inst{2}
\and 
C. Joblin \inst{1}
}
\offprints{\\ B.~Joalland, \email{joalland@cesr.fr}}
\institute{
Centre d'Etude Spatiale des Rayonnements, Universit\'e de Toulouse et CNRS, Observatoire Midi-Pyr\'en\'ees, 9 Av. du Colonel Roche, 
31028 Toulouse cedex 04, France
\and
Laboratoire de Chimie et Physique Quantiques, IRSAMC, Universit\'e de Toulouse et CNRS, 118 Route de Narbonne, 31062 Toulouse Cedex, France
}
\date{Received ?; accepted ?}

\abstract
{}
{We investigate the presence of silicon atoms adsorbed on the surface of interstellar polycyclic aromatic hydrocarbons (PAHs) to form SiPAH $\pi$-complexes.}
{We use quantum chemistry calculations to obtain structural, thermodynamic and mid-IR properties of neutral and cationic SiPAH $\pi$-complexes.}
{The binding energy was found to be at least 1.5 eV for [SiPAH]$^+$ complexes whereas it is $\sim$0.5 eV for their neutral counterparts. From the spectral analysis of the calculated IR spectra, we found that the coordination of silicon to PAH$^+$ does not strongly affect the intensities of the PAH$^+$ spectra, but systematically introduces blueshifts of the C-C in-plane and the C-H out-of-plane bands.}
{The thermodynamic data calculated for [SiPAH]$^+$ complexes show that these species are stable and can be easily formed by radiative association of Si$^{+}$ and PAH species that are known to be abundant in photodissociation regions. Their mid-IR fingerprints show features induced by the coordination of silicon that could account for (i) the blueshifted position of the 6.2 $\mu$m AIB and (ii) the presence of satellite bands observed on the blue side of the 6.2 and 11.2 $\mu$m AIBs. From such an assignment, we can deduce that typically 1\% of the cosmic silicon appears to be attached to PAHs.}

\authorrunning{Joalland et al.}
\titlerunning{[SiPAH]$^+$ in the Interstellar Medium}
\maketitle

\section{Introduction} \label{int}

\subsection{Astrophysical Context} \label{inf}

Silicon is the most depleted abundant element after iron, representing a major constituent of interstellar "standard" dust grains (\citealt{whi03}), essentially in the form of silicates. \citet{jon00} concluded from the analysis of depletion patterns and dust processes in different regions of the diffuse interstellar medium (ISM) that Si is more easily eroded from dust than any of the other elements. This is in line with the work of \citet{tie98}, who demonstrated that a significant fraction of the elemental silicon is locked up in a relatively volatile dust component with a binding energy roughly equal to 2.0 eV.

Polycyclic aromatic hydrocarbons (PAHs) that are amongst the dust components offer a large surface for atomic and molecular adsorptions. \citet{klo95} have established a correlation between metal-PAH binding energies and the observed depletion of these atoms in the ISM. This model predicts that the abundance of complexed PAH molecules can explain 5-10\% of the depletion of metallic atoms in the gas-phase of the diffuse interstellar medium. This work also emphasizes that Fe and Si are the most abundant metals bonded on PAHs, if Si is considered as a metal although it is not in strict chemical terms.

Interstellar PAHs are revealed by their emission in the aromatic infrared bands (AIBs) which set aglow the general interstellar medium (\citealt{leg84,all85}). The interpretation of the AIBs has motivated a lot of observational, experimental, and theoretical studies (see \citealt{tie08} for a recent review), but it is not yet entirely resolved, leaving the door open to search for new types of PAH-related species such as SiPAH $\pi$-complexes.

\subsection{Silicon-PAH in Physical Chemistry: Experiments and Calculations} \label{inf}

\citet{sri92} and \citet{jae05} demonstrated theoretically and experimentally that the $\pi$-system complex [SiC$_6$H$_6$]$^+$ is more stable than all the other isomers. 
Experimental studies on [SiPAH]$^+$ complexes are scarce. In the selected ion flow tube (SIFT) experiments carried in Bohme's group, it was clearly established that the capture of ground state atomic silicon ions Si$^+$ ($^2$P) by benzene and naphthalene is an efficient process in the gas phase (\citealt{boh91}). These authors also suggested a possible chemical role of the large PAH molecular surface in the catalysis of small Si-bearing molecules of astrophysical interest (\citealt{boh89}). The radiative association of Si$^+$ with a variety of PAHs was studied in the low-pressure conditions of a Fourier transform ion cyclotron resonance (FTICR) mass spectrometer in the group of Dunbar (\citealt{dunb94,poz97}). Charge transfer and condensation reactions (insertion of Si between C and H with loss of H) were found to compete efficiently with the association process. The results obtained in these low-pressure conditions can be a good indication of the processes that are likely to occur in the ISM, that is to say that [SiPAH]$^+$ complexes can be formed efficiently and play a role in the chemistry there.

Our work aims at investigating the possible relevance of [SiPAH]$^{+}$ complexes in the ISM, emphasizing two aspects: the stability of these systems and the influence of the coordination of Si on the IR spectra of PAHs. We report extensive calculations of structural, thermodynamic, and mid-infrared spectroscopic properties of ground state [SiPAH]$^{+}$ complexes for four compact PAHs of increasing size : naphthalene C$_{10}$H$_8$, pyrene C$_{16}$H$_{10}$, coronene C$_{24}$H$_{12}$ and ovalene C$_{32}$H$_{14}$. Using the methods described in section 2, we present the Si-PAH$^+$ binding energies and structures in section 3.1, and the IR calculated spectra in section 3.2 where systematic effects on band intensities and positions are also discussed. In section 4, the astrophysical implications are discussed and our main results are summarized in section 5.

\section{Methods} \label{int}

Geometry optimizations and frequency calculations were performed using well-established quantum chemical techniques in the framework of Density Functional Theory (DFT, \citealt{dft89}) with the \textsc{Gaussian03} package (\citealt{g03}).

The hybrid B3LYP functional was chosen according to its wide use in the study of the ground-state properties of PAHs and related molecules (see for instance \citealt{lan96,bau02,bau08}). The choice of the D95 basis set augmented with a complete set of diffuse and polarization functions, hereafter indicated by D95++**, was made after a comparative study with other basis sets and different combinations of diffuse and polarization functions. Geometry optimizations were systematically followed by the full vibrational analyses to confirm that the geometries are minima on the potential energy surface. These were also used to determine the discrete IR spectra in the harmonic approximation.

Errors in the calculated values of the binding energies can come from uncertainties inherent to DFT methods and the finite character of the basis set.  To give some insights into the relevance of the DFT choice and especially of the functional and the basis set choices to calculate binding energies, some test calculations were performed on the ionization potentials ($IP$s) of silicon and PAHs (cf Tab. 1). The calculated values of the $IP$s were found to be lower than the experimental ones. For PAHs, the average of this discrepancy is about -0.26 eV (from -0.20 to -0.33 eV) for vertical $IP$s. This level of calculation provides the right relative order compared to experimental data, with $IP$s systematically underestimated by 4-5\%, which is sufficiently accurate for the purpose of this work.

In the study of the vibrational properties of [SiPAH]$^+$ complexes, scaling factors have to be applied to the calculated harmonic frequencies at the DFT level for basis set incompleteness and anharmonicity effects. We determined the scaling factors for PAH$^{0}$ by comparing the calculated spectrum of each PAH$^0$ to the experimental spectrum recorded in cold matrix of argon for naphthalene (\citealt{hud98}) and in cryogenic matrix of neon for pyrene, coronene and ovalene (\citealt{job94}). We subsequently applied a scaling factor derived from the average of the scaling factors determined for each PAH to the calculated spectra of PAH$^+$ and the related [SiPAH]$^{+}$ complexes in each spectral region of interest. Three scaling factors were used, one for the 10-15 $\mu$m (1000-700 cm$^{-1}$) range, one for the 5-10 $\mu$m (2000-1000 cm$^{-1}$) range, and one for the 3 $\mu$m (3300 cm$^{-1}$) C-H stretch range where modes of larger anharmonicity are present.

\begin{table}
\caption{Computational and experimental ionization potentials ($IP$s) of the four considered PAHs and of Si. Computational values are obtained at the B3LYP/D95++** level of theory.}
\label{table1}
\begin{center}
\begin{tabular}{ccccc}

\hline \hline  
 & $IP$ (eV) & & $\Delta IP$ (eV)$^{c}$\\
 & calc.$^a$ & exp.$^{b}$ &calc. &exp.\\
 
\hline
\noalign{\smallskip}

naphthalene & 7.82/7.91 & 8.14 & -0.30/-0.21 & -0.01\\
C$_{10}$H$_8$\\
pyrene & 7.10/7.16 & 7.43 & -1.02/-0.96 & -0.72\\
C$_{16}$H$_{10}$\\
coronene & 7.02/7.09 & 7.29 & -1.10/-1.03 & -0.86\\
C$_{24}$H$_{12}$\\
ovalene & 6.40/6.43 & 6.71 & -1.72/-1.69 & -1.44\\
C$_{32}$H$_{14}$\\
Si & 8.12 & 8.15\\

\hline
\multicolumn{5}{p{8cm}}{
$^a$ adiabatic/vertical $IP$.
$^{b}$ Experimental values are extracted from the NIST database.
$^{c}$ $\Delta IP$ (eV) = $IP$(PAH) - $IP$(Si).
}
\\
\end{tabular}
\end{center}  
\end{table}

\section{Results} \label{int}

\subsection{Thermodynamics and Structures} \label{inf}

We present in this part thermodynamic and structural properties of [SiPAH]$^{+}$ complexes, completed with thermodynamic properties of model [SiPAH]$^0$ complexes. The calculated relative enthalpies at 0 K ($\Delta H$(0 K)) of the lowest energy isomers of [SiPAH]$^0$ and [SiPAH]$^+$, with respect to the most stable dissociation products (Si($^3$P)  + PAH) and (Si($^3$P)  + PAH$^+$) are reported in Tab. 2. Binding energies ($E_b$) correspond to -$\Delta H$(0 K) values. A structure is considered stable when its $E_b$ is positive. In the case of the cationic complexes, the relative enthalpies of the reaction Si$^+$($^2$P) + PAH $\to$ [SiPAH]$^+$ at 0 K ($\Delta H'$(0 K)) are also reported.

\begin{figure*}
\begin{center}
\includegraphics[height=24cm, width=17.5cm]{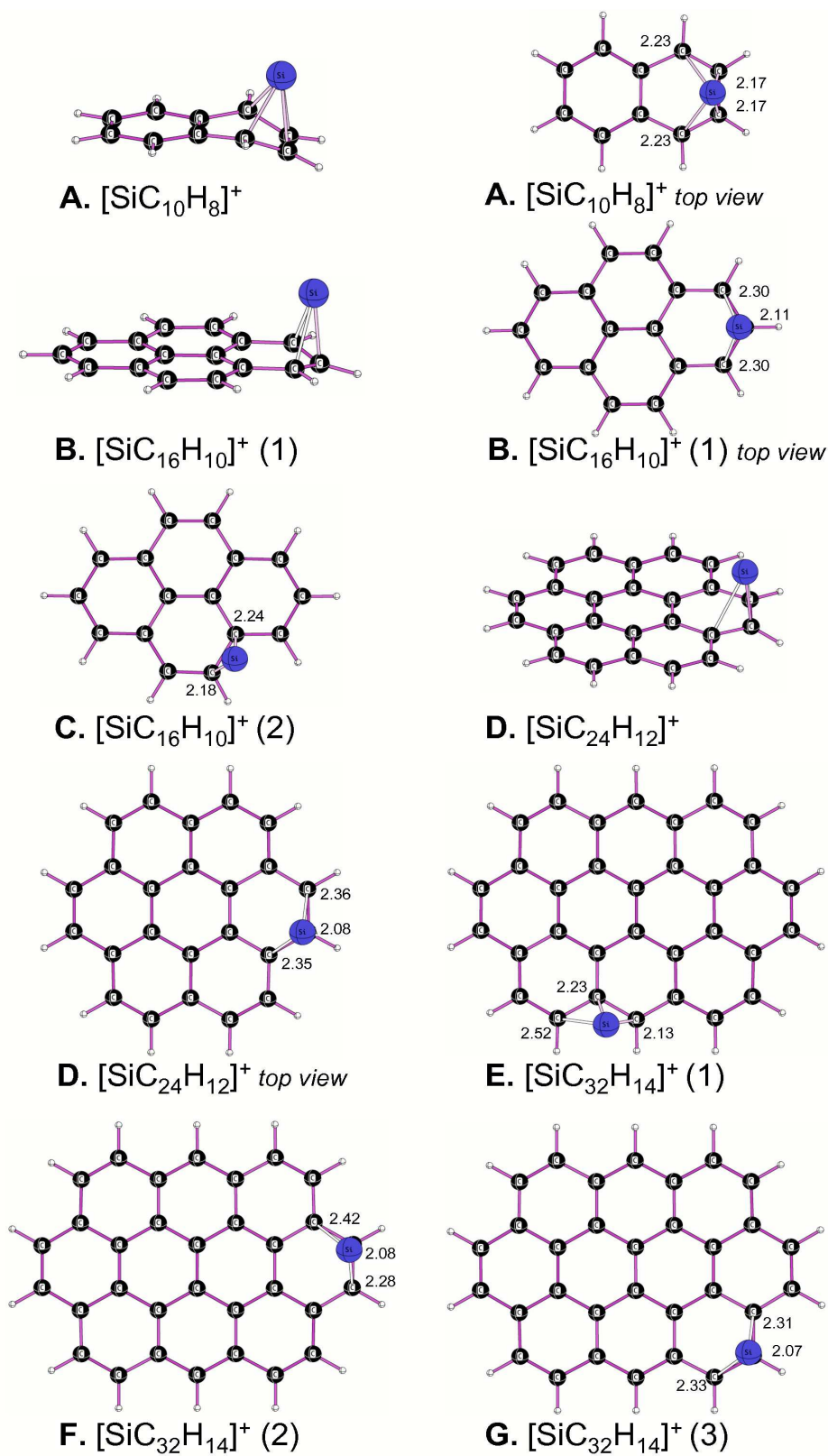}
\caption{Lowest energy stable structures for the cationic complexes of silicon with naphthalene C$_{10}$H$_8$ (structure A), pyrene C$_{16}$H$_{10}$ (structures B, C), coronene C$_{24}$H$_{12}$ (structure D), and ovalene C$_{32}$H$_{12}$ (structures E, F, G), optimized at the B3LYP/D95++** level of theory. The values expressed in \AA\ indicated on each structure correspond to the Si-C bond lengths. }
\end{center}
\end{figure*}

\begin{table*}
\caption{Electronic states and calculated relative enthalpies at 0 K of the lowest energy isomers of [SiPAH]$^{0/+}$, with respect to the most stable dissociation products (Si($^3$P) + PAH) and (Si($^3$P) + PAH$^+$) (corresponding to $\Delta H$(0 K)) and with respect to (Si$^+$($^2$P) + PAH) (corresponding to $\Delta H'$(0 K)) at the B3LYP/D95++** level of theory. $E_b$(Si-PAH$^{0/+}$) = $-\Delta H$(0 K)}
\label{table1}
\begin{center}
\begin{tabular}{ccccccc}
\hline \hline  
 & electronic & $\Delta H$(0 K) & & electronic & $\Delta H$(0 K) & $\Delta H'$(0 K)\\
 & state & (eV) & & state & (eV) & (eV)\\
\hline
\noalign{\smallskip}
$[$SiC$_{10}$H$_8]^0$ & $^3$A'' & -0.54 & $[$SiC$_{10}$H$_8]$$^+$ & $^2$A' & -2.67 & -2.99\\
$[$SiC$_{16}$H$_{10}]^0$ & $^3$A'' & -0.58 & $[$SiC$_{16}$H$_{10}]$$^+$ (1) & $^2$A & -2.08 & -3.11\\
 & & & $[$SiC$_{16}$H$_{10}]$$^+$ (2) & $^2$A & -1.86 & -2.89\\
$[$SiC$_{24}$H$_{12}]^0$ & $^3$A'' & -0.58 & $[$SiC$_{24}$H$_{12}]$$^+$ & $^2$A & -1.95 & -3.09\\
$[$SiC$_{32}$H$_{14}]^0$ & $^3$A'' & -0.49 & $[$SiC$_{32}$H$_{14}]$$^+$ (1) & $^2$A & -1.69 & -3.40\\
 & & & $[$SiC$_{32}$H$_{14}]$$^+$ (2) & $^2$A & -1.58 & -3.29\\
 & & & $[$SiC$_{32}$H$_{14}]$$^+$ (3) & $^2$A & -1.57 & -3.29\\
\hline
\end{tabular}
\end{center}  
\end{table*}

All [SiPAH]$^{0/+}$ complexes were found to have positives $E_b$, providing evidence for their thermodynamic stability. For the [SiPAH]$^0$ complexes, the $E_b$ of the most stable structures were found to lie between 0.49 and 0.58 eV, leading to an average value of $\sim$0.5 eV. They were found to be in a doublet spin-state. For the [SiPAH]$^+$ complexes, the $E_b$ of the most stable structures, found to be in a triplet spin-state, decreases from 2.67 eV to 1.69 eV as the size of the PAH increases. Interestingly, $E_b$ decreases when $\Delta IP$ increases. This is consistent with the covalent nature of a bond, assuming  the same overlap between the two orbitals in interaction. When $\Delta IP$ is equal to zero, $E_b$ is maximal. This is the case for [SiC$_{10}$H$_{8}$]$^+$ ($\Delta IP$(exp) = -0.01 eV). As the $IP$ of PAH decreases roughly exponentially with the size, approaching a value of 6.0 eV, the value of $E_b$ is expected to decrease to a value of roughly 1.5 eV for larger species. An examination of the Mulliken charges on the silicon atom and the PAH molecule shows for the larger species that the positive charge is mainly located on the PAH. This is consistent with the $\Delta IP$ values.

In the case of the formation of a [SiPAH]$^+$ molecule from a Si$^+$ and a PAH, we determined that that is an exothermic process liberating more than 3.0 eV on average.

All possible $\pi$-interaction sites of Si with PAHs were considered to find the lowest-energy isomers. Their structures are displayed in Fig. 1. We found two low-energy isomers for [SiC$_{16}$H$_{10}$]$^+$ and three low-energy isomers for [SiC$_{32}$H$_{14}$]$^+$ with very close binding energies: 2.02 and 1.86 eV for the two isomers of [SiC$_{16}$H$_{10}$]$^+$ and 1.69, 1.58, and 1.57 eV for the three isomers of [SiC$_{32}$H$_{14}$]$^+$.

In all the most stable optimized structures Si preferentially coordinates on the edge of the PAH,
mostly in a two closest neighbour C-atom configuration for neutral species and in a three closest neighbour C-atom configuration for cationic species (the so-called $\eta$$_2$ and $\eta$$_3$ edge coordination, respectively). Interestingly, the $\eta$$_2$ coordination mode was also found in the case of [FePAH]$^{0}$ complexes (\citealt{sim07}) and [CuPAH]$^+$ complexes (\citealt{dunb02}), and the $\eta$$_3$ coordination mode in the case of [SiC$_6$H$_6$]$^+$ complex (\citealt{jae05}). These coordination modes imply both a curvature of the edge of the C-skeleton and ring distortions which had been already observed in the case of [TM-C$_6$H$_6$]$^+$ (TM: transition metals) calculated structures  (\citealt{jae04}).

Besides the fact that silicon is mainly present in its ionized form in the gas-phase of photodissociation regions (PDRs) where PAH$^{0/+}$ are largely detected, the cationic complexed species are more interesting to study for their higher stability. Therefore only the spectral features of the latter species and their astrophysical implications are discussed in the following sections.

\subsection{Influence of the Coordination of Silicon on the IR Spectra of PAH$^+$} \label{inf}

As mentioned in Sect. 1, much work has been devoted to the identification of the carriers of the astronomical AIBs. The main bands are located at 3.3, 6.2, 7.7, 8.6, 11.2 and 12.7 $\mu$m. The 3.3 $\mu$m band is attributed to the C-H stretch ($\nu$$_{C-H}$), the 6.2 $\mu$m band to pure C-C stretches ($\nu$$_{C-C}$), the 7.7 $\mu$m band to a mixture of C-C stretches and in-plane C-H bends, the 8.6 $\mu$m band to pure in-plane C-H bends ($\delta$$_{C-H}$). The 11.2 and 12.7 $\mu$m are both assigned to out-of-plane (oop) C-H bends ($\gamma$$_{C-H}$). The positions of these $\gamma$$_{C-H}$ reflect the number of adjacent C-H groups on the edge of the PAH C-skeleton. A classification was established by \citet{hon01}. The C-H groups carried by aromatic rings with no other C-H group are called solo C-H groups and lead to IR activity around the 11.2 $\mu$m AIB. Likewise, C-H groups with one and two adjacent C-H groups are respectively called duo and trio C-H groups and lead to IR activity in the range of the 12.7 $\mu$m AIB.

The IR calculated spectra of all stable isomers of [SiPAH]$^+$ complexes are reported in Fig. 2 in the [3-15] $\mu$m range. The spectrum of the corresponding bare PAH$^+$ is reported at the bottom of each figure to make comparisons easier. After calibration the calculated discrete spectra are convoluted by a Lorentzian line profile with an assumed FWHM (full width at half maximum) of 0.2 $\mu$m, which accounts reasonably well for the band broadening of the astronomical spectra. The $I_{6-10\mu m}/I_{10-15\mu m}$ intensity ratios which are often used to analyze observed spectra are presented in Table 3, along with the relevant shifts occuring for the $\gamma$$_{C-H}$ and $\nu$$_{C-C}$ bands. We emphasize that the extracted shifts induced by the coordination of silicon to the PAH depend only marginally on the spectral calibration (a maximal deviation of -0.02 $\mu$m without applying scaling factors).
 
\begin{figure*}
\begin{center}
\includegraphics[height=24cm, width=17cm]{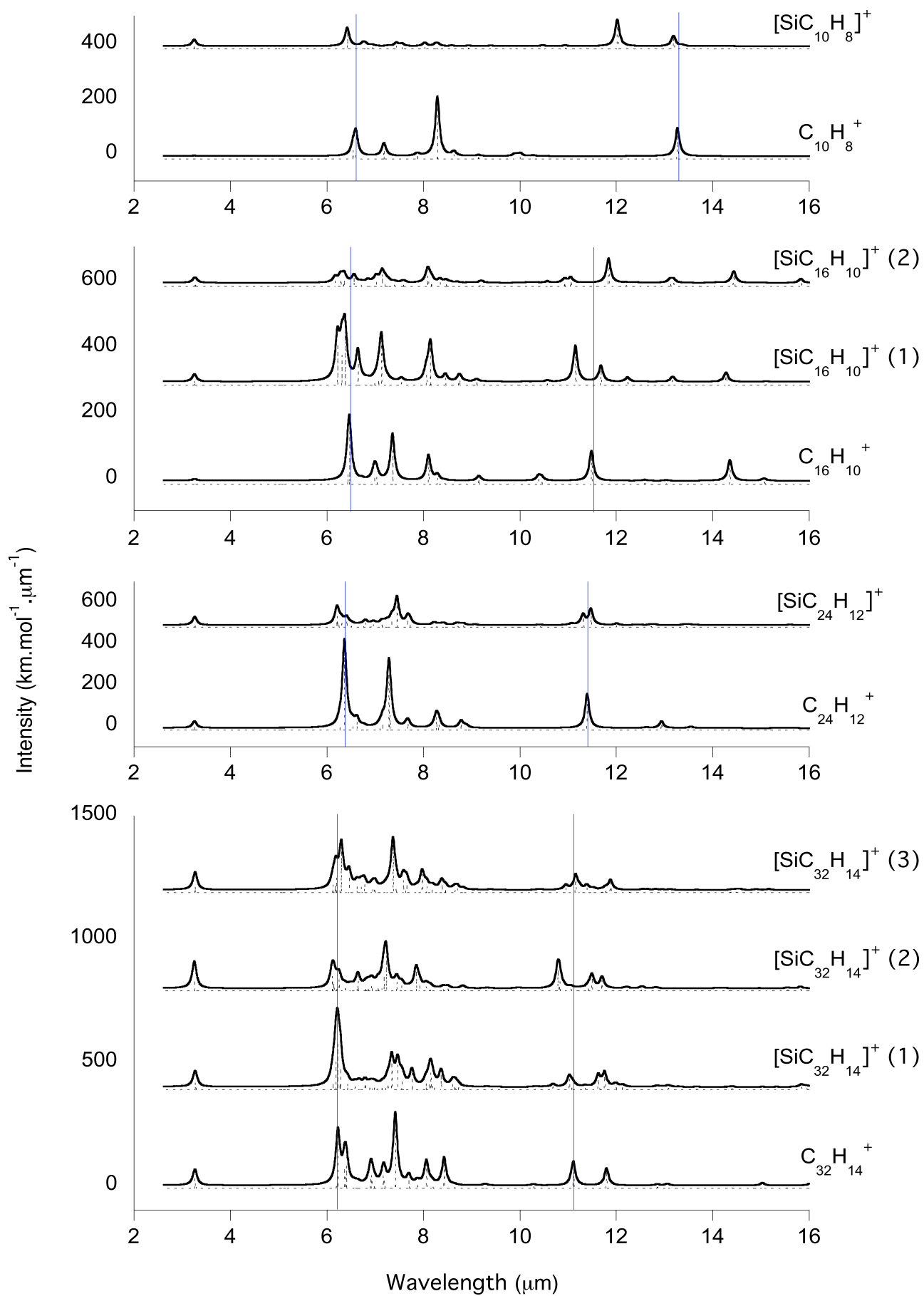}
\vspace{-0.0cm}
\caption{IR spectra of the most stable structures of PAH$^{+}$ and their related Si-complexes calculated at the B3LYP/D95++** level of theory. Discrete spectra are convoluted by a Lorentzian profile with a FWHM of 0.2 $\mu$m.
The derived average scaling factors are 0.953 for the 3-5 $\mu$m range, 0.976 for the 5-10 $\mu$m range, and 0.966 for the 10-15 $\mu$m range.}
\end{center}
\label{ext}
\vspace{-0.5cm}
\end{figure*}

\begin{table*}
\caption{$I_{6-10\mu m}/I_{10-15\mu m}$ ratios ($R$) for the ground states of PAH$^+$ and their complexes with Si and the effects of the complexation on this ratio.}
\label{table1}
\begin{center}
\begin{tabular}{ccccccc}
\hline \hline  
 & & $R$(SiPAH$^+$)/ & & $\delta$$\lambda$ oop C-H bend $^*$ & &  $\delta$$\lambda$ ip C-C stretch $^*$\\
 & $R$ & $R$(PAH$^+$) & $\lambda$ oop C-H bend & SiPAH$^+$ vs PAH$^+$ &  $\lambda$ ip C-C stretch $^*$ & SiPAH$^+$ vs PAH$^+$ $^*$\\
\hline
\noalign{\smallskip}
C$_{10}$H$_8$$^+$ & 2.92 & & 13.28 & & 6.60 (6.55) & \\
$[$SiC$_{10}$H$_8]^+$ & 0.97 & 0.33 & 13.19 & -0.09 & 6.43 & -0.17\\ 
 & & & 12.02 & new & & \\
C$_{16}$H$_{10}$$^+$ & 2.86 & & 11.46 & & 6.47 & \\
$[$SiC$_{16}$H$_{10}]$$^+$ (1) & 3.56 & 1.24 & 11.68 & +0.22 & 6.22 & -0.25\\ 
 & & & 11.15 & -0.31 ; new & & \\
$[$SiC$_{16}$H$_{10}]$$^+$ (2) & 1.58 & 0.55 & 11.85 & +0.37 & 6.18 & -0.29\\ 
C$_{24}$H$_{12}$$^+$ & 5.24 & & 11.41 & & 6.38 & \\
$[$SiC$_{24}$H$_{12}]$$^+$ & 2.87 & 0.55 & 11.48 & +0.07 & 6.21 & -0.17\\
 & & & 11.32 & -0.09 ; new & & \\
C$_{32}$H$_{14}$$^+$ & 7.27 & & 11.11 & & 6.23 & \\
$[$SiC$_{32}$H$_{14}]$$^+$ (1) & 5.01 & 0.69 & 11.03 & -0.08 & 6.13 & -0.10\\
$[$SiC$_{32}$H$_{14}]$$^+$ (2) & 6.63 & 0.91 & 11.49 & +0.38 & 6.11 & -0.12\\
 & & & 10.79 & -0.33 ; new & & \\
$[$SiC$_{32}$H$_{14}]$$^+$ (3) & 5.34 & 0.73 & 11.15 & +0.04 & 6.13 & -0.10\\
 & & & 10.95 & -0.16 ; new & & \\
\hline
\multicolumn{5}{p{8cm}}{
$^*$ All values are expressed in $\mu$m.}
\\
\end{tabular}
\end{center}  
\end{table*}

As can be seen in Fig. 2, the global features of the IR spectra of [SiPAH]$^+$ complexes remain close to those of bare PAH$^+$. This contrasts with the effect of the coordination of Fe to PAH$^+$ (\citealt{hud05,szc06,sim07,sim08a}) and can be understood considering that the charge remains mainly located on the PAH with respect to the values of the IPs of Si and PAH. 
In particular, the 3.3 $\mu$m band is hardly affected by the coordination of silicon, in either position or intensity.

For the small [SiC$_{10}$H$_8$]$^+$ complex, the $I_{6-10\mu m}/I_{10-15\mu m}$ ratio decreases substantially with the coordination of silicon (2.92 to 0.92) whereas for the largest species, the $I_{6-10\mu m}/I_{10-15\mu m}$ ratios remain of the same order of magnitude (2.86 to 2.57 for pyrene, 5.24 to 2.87 for coronene, 7.27 to 5.66 for ovalene; taking the average value for several isomers). A simple explanation in terms of partial charge transfer seems sufficient to account for this trend. Given the fact that PAH$^+$ molecules show strong features in the [6-10] $\mu$m range while PAH$^0$ do not (\citealt{lan96}), the observed decrease of the $I_{6-10\mu m}/I_{10-15\mu m}$ ratio with respect to the one of bare PAH$^+$ is due to the coordination of silicon which reduces the charge carried by the PAH molecule. When the size of the PAH grows, the charge remains mainly located on the PAH with respect to the values of their IPs, therefore the coordination of silicon has a moderated effect on the total IR activity in this region, in particular for the largest species. This differs with the effect of the coordination of Fe on PAH$^+$ which leads to a strong decrease of the $I_{6-10\mu m}/I_{10-15\mu m}$ ratio (\citealt{sim07,sim08a}). 

In contrast, the coordination of silicon leads to significant wavelength shifts. Several new modes are activated, mostly intense in the [6-8] $\mu$m range, while the activity in the [8-10] $\mu$m is reduced. All the corresponding vibrations in this region are initially in-plane features before complexation. Since the coordination of silicon on PAH surface breaks the plane of the PAH molecule, that leads to a scattering of the molecule vibrations in many coupled modes. We point out the appearance of strong IR activities in the 6.0 $\mu$m range, due to $\nu$$_{C-C}$ modes with C atoms bounded to Si. Then [SiPAH]$^+$ show some intense $\nu$$_{C-C}$ bands while introducing an IR activity at shorter wavelengths than the band of the bare PAH$^+$ with an apparent blueshift average value of -0.17 $\mu$m. However, the value of the shift seems to decrease when the size of the complexed PAH molecule increases, consistently with the reason that induces this shift: the larger the PAH is, the less significant the break of its plane is. The calculated value of -0.17 $\mu$m has therefore to be taken with caution.

In the [10-15] $\mu$m range of $\gamma$$_{C-H}$ modes, the coordination of silicon leads to a  splitting of the bands, but the total IR activity is preserved. Even though the $\gamma$$_{C-H}$ band positions depend on the number of adjacent hydrogens which varies from four in naphthalene to one in ovalene, complex formation increases the IR activity in the shortest wavelength domain. Indeed, the coordination of silicon causes a splitting of the main $\gamma$$_{C-H}$ band of PAH$^+$ in two bands: one at lower, and the other one at larger wavelength with respect to those of PAH$^+$. For instance, the bands at lower and larger wavelengths shift respectively towards 11.43 and 11.97 $\mu$m for [SiC$_{16}$H$_{10}$]$^+$ (1), while the corresponding band of C$_{16}$H$_{10}$$^+$ is located at 11.76 $\mu$m. The band at lower wavelengths, located in the solo domain, is due to the $\gamma$$_{C-H}$ mode of the C-H bond with its motion hindered by the presence of Si. An interesting feature is that this band is intense and occurs for all [SiPAH]$^+$ complexes with Si coordinated above a C atom bounded to a H atom (cf Fig. 1, [SiC$_{10}$H$_{8}$]$^+$, [SiC$_{16}$H$_{10}$]$^+$ (1), [SiC$_{24}$H$_{12}$]$^+$, [SiC$_{32}$H$_{14}$]$^+$ (2) \& (3)). The average shift of this band position with respect to the position of the main $\gamma$$_{C-H}$ band of the corresponding bare PAH$^+$ is -0.22 $\mu$m. Contrary to the shift observed in the 6.0 $\mu$m range, this phenomenon of splitting does not depend on the molecular size but only on the geometry: the coordination of Si above a C atom bounded to a H systematically hinders the C-H vibration. For all these isomers, this band is the most intense when its intensity is normalized to the number of involved H atoms (cf Fig. 3). \citet{job94} pointed out from experimental spectra on neutral PAHs the four time stronger intensity of solo $\gamma$$_{C-H}$ than other $\gamma$$_{C-H}$, which is in line with our calculations. The strong intensity and the position of the new $\gamma$$_{C-H}$ induced by the coordination of silicon are consistent with a solo character.

\begin{figure}[!]
\begin{center}
\includegraphics[height=8cm, width=9cm]{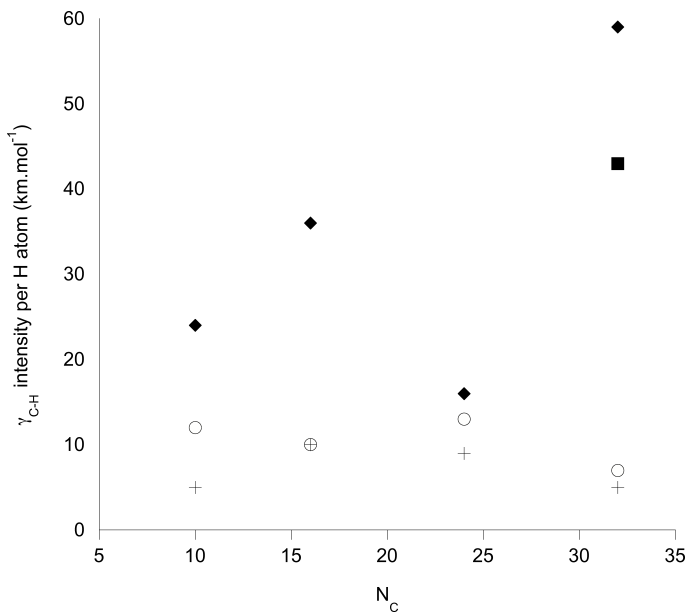}
\caption{Comparison between the $\gamma$$_{C-H}$ intensity per H atom for PAH$^+$ (rings) and [SiPAH]$^+$ (diamonds for $\gamma$$_{C-H}$ additional band induced by the coordination of Si and crosses for the remaining $\gamma$$_{C-H}$, strongest band in case of isomers) considered as a function of the system size N$_C$. The square represents the solo $\gamma$$_{C-H}$ intensity per H atom for C$_{32}$H$_{14}$$^+$.
}
\end{center}
\end{figure}

\begin{figure*}
\begin{center}
\includegraphics[height=6cm, width=8cm]{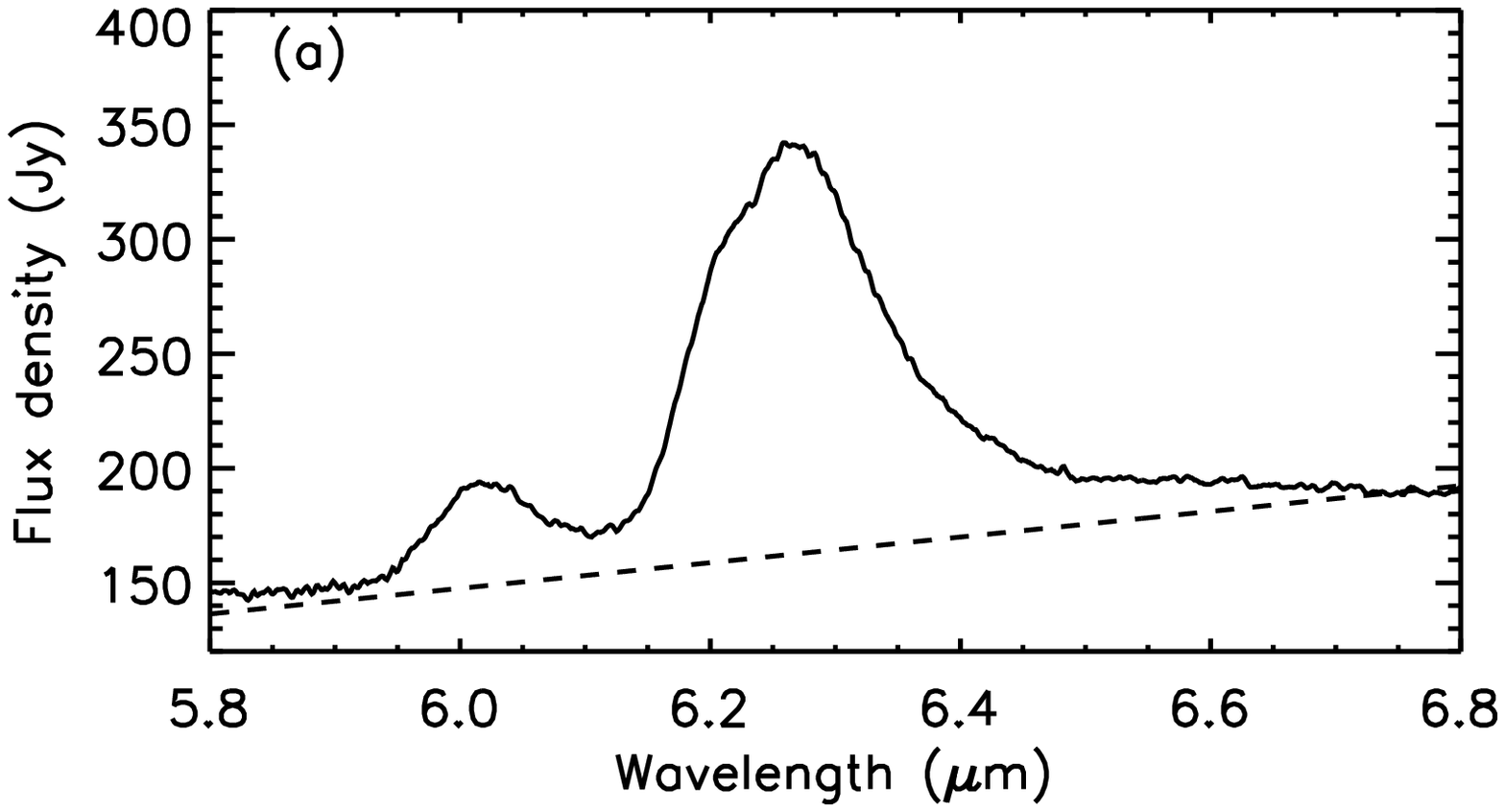}
\includegraphics[height=6cm, width=8cm]{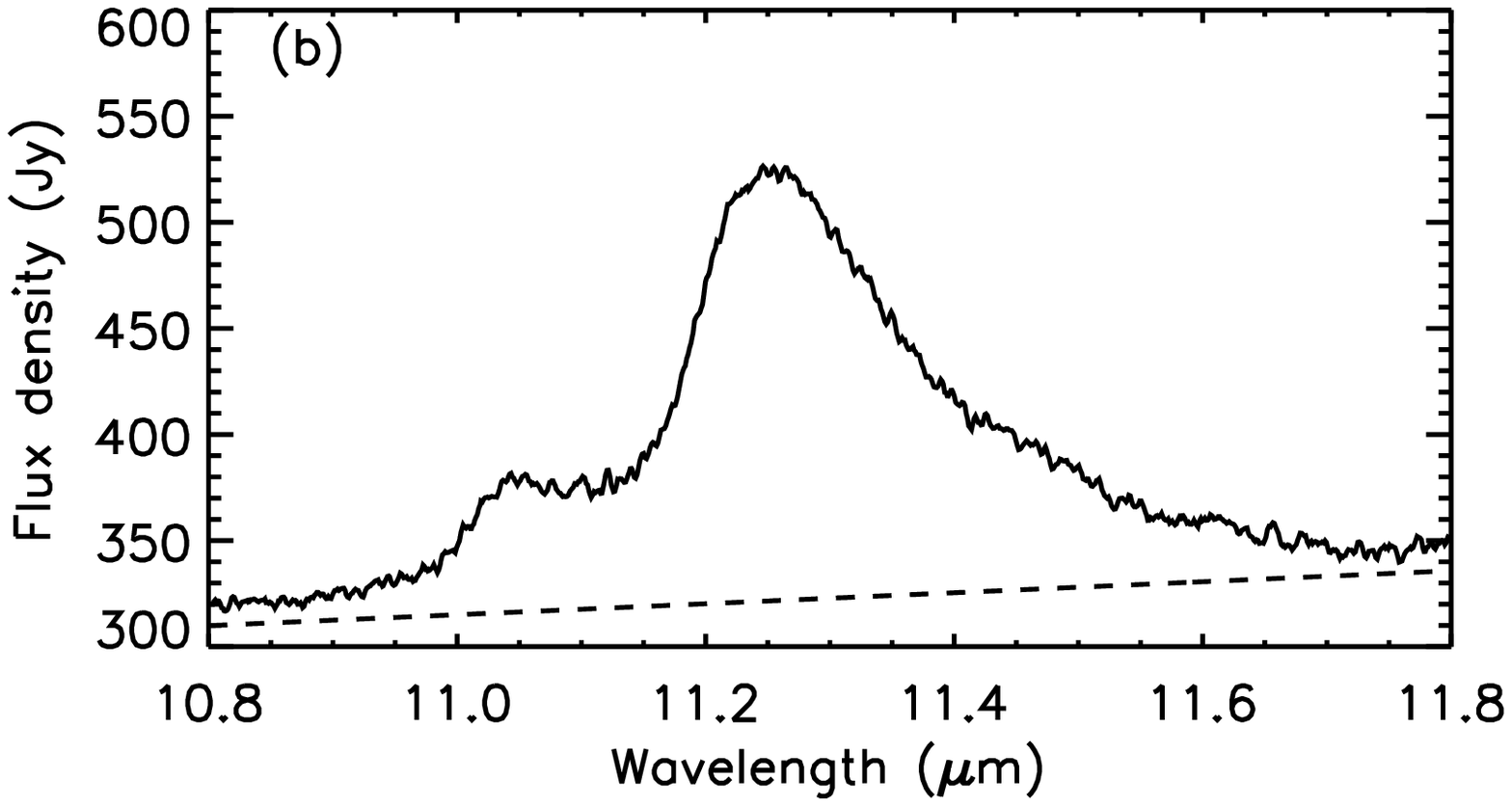}
\caption{The IR spectrum of the Red Rectangle nebula retrieved from the ISO archive ; (a) the [5.8-6.8] $\mu$m range and (b) the [10.8-11.8] $\mu$m range with their underlying continua (dotted line).}
\end{center}
\end{figure*}

\section{Astrophysical Implications} \label{int}

Our thermodynamic results show that the binding energies of [SiPAH]$^+$ complexes are at least 1.5 eV.  Since they are three times larger than those of [SiPAH]$^0$, [SiPAH]$^+$ complexes are more likely to survive in the conditions of the ISM than their neutral counterparts. Furthermore the formation of [SiPAH]$^+$ complexes by radiative association is an exothermic process of 3.0 eV without any expected activation barrier (cf Tab. 2, calculated $\Delta H'$(0 K)). The formation of [SiPAH]$^+$ could therefore occur in many regions where Si$^{+}$ and PAH$^{0}$ are abundant, as in PDRs (see for instance \citealt{kau06}). These [SiPAH]$^+$ complexes could then represent a significant fraction of PAH-related molecules in such regions.

The spectral analysis of the calculated spectra of [SiPAH]$^+$ provides insights for astronomical detection of such species. Indeed, the coordination of silicon to the PAH$^+$ surface has two main effects:

1. a strong IR activity at the blueshifted side of the bare PAH$^+$ $\nu$$_{C-C}$ band. The corresponding shift is going from -0.30 down to -0.10 $\mu$m, and seems to be decreasing when the size of the PAH increases.

2. the occurrence of an additional $\gamma$$_{C-H}$ band for many isomers, blueshifted with respect to the lowest wavelength $\gamma$$_{C-H}$ band of PAH$^+$. The average shift is -0.2 $\mu$m, independent on the size of the complex but characteristic of the coordination geometry.

Following the profile variability of the 6.2 $\mu$m AIB from source to source, \citet{pee02} pointed up two main components at 6.2 and 6.3 $\mu$m - classes A and B. This has motivated further experimental and theoretical studies. In particular \citet{hud05} computationally established that the class B can be explained by the emission of large PAH$^+$, while the assignment of the class A feature is not that clear: the effect of the substitution of one or more C atoms by hetero-atoms (N, O, Si) and the coordination of metal ion (Fe$^+$, Mg$^+$, Mg$^{2+}$) to PAH was investigated. That study puts forward the role of the insertion of N atoms within the C-skeleton to resolve this question. On the other hand, \citet{pin08} recently showed on the basis of IR spectroscopy of soot-like material that the position of the 6.2 $\mu$m AIB can be explained by an evolutionary scenario of carbon dust through its aromatic to aliphatic content. This proposal awaits the right analogues in the laboratory of the "astro-PAHs", which have to be small enough to respect the stochastic heating constraint.

The $\nu$$_{C-C}$ shift induced by the coordination of Si to PAH$^+$ that we calculated is at least -0.1 $\mu$m. For larger PAHs, we expect a similar shift due to the preference of Si to coordinate on the edge of the PAH, whatever its size, and then its abilility to break its plane. Therefore this value can be applied to interstellar PAHs which are likely larger than the species we studied (\citealt{leg84,bau08}). Our study provides an alternative scenario to the polycyclic aromatic nitrogen heterocycles (PANHs) explanation and the aliphatic to aromatic evolution scheme for the shortest wavelength position of 6.2 $\mu$m AIB.

Considering that the 11.2 $\mu$m AIB is the signature of the $\gamma$$_{C-H}$ band of interstellar PAHs, we could predict a band located at 11.0 $\mu$m if [SiPAH]$^+$ complexes are present. Interestingly, such a band has been detected in many astronomical objects by the Short Wavelength Spectrometer (SWS) on board the Infrared Space Observatory (ISO) (\citealt{hon01,pee02,die04}). We also note that an additional band at 6.0 $\mu$m of comparable intensity is often present. The presence of [SiPAH]$^+$ complexes could account for this additional feature: we have seen that the $\nu$$_{C-C}$ shift is larger for small [SiPAH]$^+$ and therefore these species may be the carriers of this satellite AIB. Another possibility could be larger complexes containing several Si atoms. We expect the shift to increase with the number of Si atoms although we have not theoretically explored this effect yet.

Fig. 4 shows the IR spectrum of the Red Rectangle, which is the brightest AIB spectrum in the sky. This region is known to be a peculiar C-rich nebula, where the low metal abundances are expected to be linked to depletion onto dust (\citealt{wae96}). In this object, the additional band at 6.0 $\mu$m, the structure in the blue wing of the 6.2 $\mu$m AIB and the 11.0 $\mu$m band are clearly visible on the spectrum retrieved from the ISO archive. The strong 7.8 $\mu$m band observed in this object implies that large ionized PAHs exist (\citealt{bau08,job08}). The latter can offer a large surface for multiple Si atoms adsorption, which strengths the possible assignment of the 6.0 $\mu$m band to the resulting complexes.

The region of the 6.2 $\mu$m AIB was fitted by three Lorentzian profiles centered at 6.0, 6.2 and 6.3 $\mu$m after a linear continuum substraction. The integrated intensities of the 6.0 and 6.2 $\mu$m bands were found to represent 13\% and 22\% of the total intensity respectively. Considering that the $\nu$$_{C-C}$ intensity is not reduced by the coordination, 35\% of PAHs would be involved in such species. Taking into account that the relative abundance of C in the framework of interstellar PAHs [C$_{PAH}$]/[H] is equal to 6.5 $\times$ 10$^{-5}$ (\citealt{job92}) and that PAHs are constituted by an average of 80 C atoms, that leads to a relative abundance [PAH]/[H] = 8 $\times$ 10$^{-7}$ in agreement with the value given by \citet{kau06}. With regard to the cosmic abundance of Si ([Si]/[H] = 3.2 $\times$ 10$^{-5}$ from \citealt{asp05}), we find that the silicon abundance which would be involved in these species corresponds to an order of magnitude of 1\%.

We compared the areas between the 11.0 and the 11.2 $\mu$m bands using the same method. The area of the 11.0 $\mu$m band represents 6\% of the total area composed of the 11.0 and 11.2 $\mu$m bands. This value seems to be consistent with the higher value extracted at 6.2 $\mu$m taking into account that (i) a redshifted counterpart which falls in the wing of the intense 11.2 $\mu$m AIB is expected at $\sim$11.4 $\mu$m and (ii) the coordination geometry does not lead to a systematic activation of the 11.0 $\mu$m vibration mode.

 \section{Conclusions} \label{int}

\citet{tie98} proposed that some silicon is locked up in a dust component which differs from the silicates and whose binding energy $E_b$ is $\sim$2 eV. Our thermodynamic results show that [SiPAH]$^+$ complexes are good candidates with a minimum binding energy $E_b$ of 1.5 eV. The formation of such species in the ISM is expected to occur by radiative association  of Si$^{+}$ with a PAH$^{0}$ molecule, which is an exothermic process liberating 3.0 eV.

The comparison of the $I_{6-10\mu m}/I_{10-15\mu m}$ ratio for PAH$^{+}$ and their Si-complexes illustrates the charge effect induced by the coordination of silicon: this does not strongly affect the IR intensities of the spectra of PAH$^+$. Hence the nature of silicon enables the formation of cationic complexes in which the charge is mainly located on PAHs. That leads the [SiPAH]$^+$ complexes to show IR spectra close to those of PAH$^+$ in terms of intensities, along with characteristic features induced by the coordination of silicon.

We propose that the presence of [SiPAH]$^+$ species in astronomical environments can be deduced from the 11.0 $\mu$m satellite band. It can also provide an explanation for the blueshifted position of the 6.2 $\mu$m AIB. Additionally, the 6.0 $\mu$m satellite band could be the signature of multiple Si atoms complexes. A firm assignment would require further experimental and theoretical work. We suggest that this identification does not require stringent constraints on the silicon abundance since typically 1\% of the cosmic silicon would be attached to PAHs.

A previous theoretical study on [FePAH]$^+$ species has not revealed specific fingerprints that could be used to identify these species in the ISM (\citealt{sim07}). Our results on [SiPAH]$^+$ provide the first spectroscopic evidence of the presence of metal-PAH species in interstellar and circumstellar environments.

\begin{acknowledgements}
We acknowledge the anonymous referee for his fruitful comments on the manuscript, and Paolo Pilleri for his help with the ISO spectrum.
This work was supported by French National Program 'Physique et Chimie du Milieu Interstellaire' which is gratefully acknowledged. We also thank the CALMIP supercomputing facility of Universit\'e de Toulouse where DFT calculations were performed.

\end{acknowledgements}

\bibliographystyle{aa}  
\bibliography{biblio}

\end{document}